\documentstyle[prd,aps,preprint,psfig]{revtex}

\newcommand{\D}{ {\rm D} }
\newcommand{\T}{ {\rm T} }
\newcommand{\hethree}{{\rm ^3He}}
\newcommand{\hefour}{{\rm ^4He}}
\newcommand{\lisix}{{\rm ^6Li}}
\newcommand{\liseven}{{\rm ^7Li}}
\newcommand{\ev}{ {\rm eV} }

\newcommand{\mev}{ {\rm MeV} }
\newcommand{\gev}{ {\rm GeV} }
\newcommand{\tev}{ {\rm TeV} }

\newcommand{\order}{{\cal O}}

\begin{document}
\tighten

\preprint{
\noindent
\begin{minipage}[t]{3in}
\begin{flushright}
TU-608 \\
YITP-00-73\\
hep-ph/0012279 \\
December 2000 \\
\end{flushright}
\end{minipage}
}

\title{Radiative decay of a massive particle and the non-thermal
process in primordial nucleosynthesis}
\author{Masahiro Kawasaki}
\address{Research Center for the Early
Universe, School of Science, University of Tokyo, Tokyo 113-0033,
Japan}
\author{Kazunori Kohri}
\address{Yukawa Institute for Theoretical Physics, Kyoto University,
Kyoto, 606-8502, Japan}
\author{Takeo Moroi}
\address{Department of Physics, Tohoku University, Sendai 980-8578, Japan}

\maketitle
\begin{abstract}

     We consider the effects on big bang nucleosynthesis (BBN) of the
     radiative decay of a long-lived massive particle.  If high-energy
     photons are emitted after the BBN epoch ($t \sim 1 - 10^3$ sec),
     they may change the abundances of the light elements through
     photodissociation processes, which may result in a significant
     discrepancy between standard BBN and observation.  Taking into
     account recent observational and theoretical developments in this
     field, we revise our previous study constraining the abundance of
     the radiatively-decaying particles.  In particular, on the
     theoretical side, it was recently claimed that the non-thermal
     production of $^6$Li, which is caused by the photodissociation of
     $\hefour$, most severely constrains the abundance of the
     radiatively-decaying particle.  We will see, however, it is
     premature to emphasize the importance of the non-thermal
     production of $^6$Li because (i) the theoretical computation of
     the $^6$Li abundance has large uncertainty due to the lack of the
     precise understanding of the $^6$Li production cross section, and
     (ii) the observational data of $^6$Li abundance has large errors.

\end{abstract}



\section{Introduction}

Big-bang nucleosynthesis (BBN) is one of the most important tools to
probe the early universe. Because it is very sensitive to the
condition of the universe from $10^{-2}$ sec to $10^{12}$ sec, we can
indirectly check the history of the universe and impose constraints on
a hypothetical particles by the observational light element
abundances.

There are a lot of models of the modern particle physics beyond the
standard model, e.g. supergravity or superstring theory, which predict
unstable massive particles with masses of $\order$(100 GeV), such as
gravitino, Polonyi field, moduli, and so on.  They have long lifetimes
because their interactions are suppressed by inverse powers of the
gravitational scale.  Consequently, these exotic particles may decay
at about the BBN epoch ($T \lesssim 1$MeV).  If the massive particles
radiatively decays, the emitted high energy photons induce the
electromagnetic cascade process.  If the decay occurs after the BBN
starts, the light elements would be destroyed by the cascade photons
and their abundances would be changed significantly.  Comparing the
theoretically predicted light element abundances with the
observational ones, we can impose constrains on the energy density,
the mass, and the lifetime of the parent massive
particle~\cite{EGLNS,KM,HKKM}.  In particular, Holtmann and the
present authors~\cite{HKKM} performed the Maximum Likelihood analysis
including both theoretical and observational errors and obtained the
precise constraints.

After Ref.~\cite{HKKM} was published, several new observational data
of light elements were reported.  As for the $^4$He abundance, it was
still unclear whether the observational value of the primordial $^4$He
mass fraction $Y$ is low ($\sim$ 0.234)~\cite{pagel,OliSkiSte} or high
($\sim$ 0.244)~\cite{Izo}.  However, Fields and Olive considered the
HeI absorption effect and reanalyzed the data~\cite{FieOLi}, and they
obtained a relatively middle value of $Y$ ($ \sim 0.238$).  On the
other hand, as for the primordial D/H, although low values of D/H
($\sim 10^{-5}$)~\cite{BurTyt} had been measured and regarded as the
primordial abundance, a relatively high value of D/H ($\sim 10^{-4}$)
was claimed again by Tytler et al.\ in the high redshift QSO
absorption systems~\cite{tyt_high}.  In their paper they stressed that
while the data may be inadequate to definitely conclude it to be
precise value, there is still a possibility of the high D/H.

On the theoretical side, recently it was claimed that the severest
constraint on the radiatively-decaying particle may be from the
non-thermal production of $^6$Li, which is a secondary $^6$Li
production due to the background $\hefour$ and the energetic T or
$\hethree$ produced by the $\hefour$ photodissociation.\footnote{
Such a possibility of the secondary process had already been pointed
out by the earlier works for hadronic decaying particles~\cite{DEHS}.}
However, the observational data of the primordial component of $^6$Li
has large uncertainties.  In addition, precise experimental data for
the nuclear cross sections are not available.  Therefore, it is
unclear how important the non-thermal $^6$Li production is once we
take account of these uncertainties.

With these new developments in theory and observation, we revise the
previous constraint on the radiative decay of long-lived particles.
We obtain the photon spectrum by solving the Boltzmann equation
numerically~\cite{KM}.  In addition we perform the Monte Carlo
simulation which includes both the experimental and theoretical
errors.  Then, we estimate the confidence levels by performing the
Maximum Likelihood method including both the theoretical and the
observational errors.

This paper is organized as follows. In Sec.~II we briefly review the
current status of the observational data. In Sec.~III we introduce the
formulations for the photodissociation and non-thermal $^6$Li
production. In Sec.~IV we compare the theoretical predictions with the
observations. Sec~V is devoted to conclusions.

\section{Observational light element abundances}
\label{sec:obs_status}

Here we summarize the current status of the observational light
element abundances.  The primordial D/H is measured in the high
redshift QSO absorption systems.  Recently a new deuterium data was
obtained from observation of QSO HS 0105+1619 at z =
2.536~\cite{ap0011179}.  It was found that the cloud is neutral and
has a simple structure.  Five Lyman series transitions caused by D and
H were observed there.  The reported value of the deuterium abundance
was relatively low, (D/H)$^{obs} = ( 2.54 \pm 0.23 ) \times 10^{-5}$.
Combined with the previous ``Low D'' data which were obtained by the
clouds at z = 3.572 towards Q1937-1009 and at z = 2.504 towards
Q1009+2956~\cite{BurTyt}, the primordial abundance is obtained as
\begin{equation}
      \label{lowd}
     {\rm Low D} : \left( {\rm D/H} \right)^{obs}
     = (3.0 \pm 0.4) \times 10^{-5}.
\end{equation}
We call this value ``Low D.'' On the other hand, Webb et al.\ oobserved
high deuterium abundance in relatively low redshift absorption systems
at z = 0.701 towards Q1718+4807~\cite{webb},
\begin{equation}
      \label{eq:highd}
      {\rm High D} : \left( {\rm D/H} \right)^{obs} = (2.0 \pm 0.5)
      \times 10^{-4}.
\end{equation}
Tytler et al.\ \cite{tyt_high} also observed the clouds independently
and obtained the similar value.  Since Webb et al.\ and Tytler et al.\ 
did not obtain the full spectra of the Lyman series in their
observations, the precise fitting of D/H based on the ``High D'' data
might be inadequate.  However, the possibility of ``High D'' have not
been excluded yet.  Therefore, we also consider the possibility of
``High D'' and include it in our analysis.

For $^3$He, we use the pre-solar measurements.  In this paper, we do
not rely upon models of galactic and stellar chemical evolution
because of the large uncertainty in extrapolating back to the
primordial abundance. But it is reasonable to assume that $^3$He/D is
an increasing function of the cosmic time, because D is the most
fragile isotope and is always destroyed whenever $^3$He is
destroyed. Using the solar-system data reanalyzed by
Geiss~\cite{Geiss93},
\begin{equation}
      \label{geiss}
      r_{3,2,\odot}^{obs} \equiv
      \left( \hethree / \D \right)^{obs}_{\odot} = 0.591 \pm 0.536,
   \label{He3/D}
\end{equation}
where $\odot$ denotes the pre-solar abundance.
We take this to be an upper bound on the primordial $^3$He to D ratio
$r_{3,2}^{obs}$:
\begin{equation}
      \label{upper_bound32}
      r_{3,2}^{obs} \le r_{3,2,\odot}^{obs}.
\end{equation}
Although in the standard scenario the theoretical prediction satisfies
the above constraint, $^4$He photodissociation produces both D and
$^3$He and can raise the $^3$He to D ratio~\cite{Sigl}.  Hence, we
include this constraint into our analysis.

The primordial $^4$He mass fraction $Y$ is inferred from observation
of recombination lines from the low metallicity extragalactic HII
regions.  Since $^4$He is produced with Oxygen in stars, the
primordial value is obtained to regress to the zero metallicity O/H
$\rightarrow 0$ for the observational data.  Recently, Fields and
Olive~\cite{FieOLi} reanalyzed the data including the HeI
absorption effect and they obtained
\begin{equation}
      \label{FieOLi}
       Y^{obs} = 0.238 \pm (0.002)_{stat} \pm (0.005)_{syst},
\end{equation}
where the first error is the statistical uncertainty and the second
error is the systematic one. We adopt the above value as the
observational  $Y$.

The primordial $^7$Li/H is observed in the Pop II old halo stars.  We
adopt the recent measurements by Bonifacio and Molaro~\cite{BonMol}
\begin{equation}
      \label{li7}
      {\rm log_{10}}\left[ \left(\liseven/{\rm H}\right)^{obs} \right]
      =-9.76 \pm (0.012)_{stat} \pm (0.05)_{syst} \pm (0.3)_{add}.
\end{equation}
Here we have added the additional uncertainty for fear that $\liseven$
in halo stars might have been supplemented (by production in
cosmic-ray interactions) or depleted (in stars)~\cite{factor-of-two}.

It is much more difficult to observe the primordial component of
$^6$Li because $^6$Li is so much rarer than $^7$Li.  Unfortunately,
enough data have not been obtained to find the ``Spite plateau'' of
$^6$Li.  However, we can set an upper bound on $^6$Li/$^7$Li, since it
is generally believed that the evolution of $^6$Li is dominated by the
production through p,$\alpha$-C,N,O cosmic ray spallation (reactions
of cosmic rays with the interstellar medium).  Intrinsically the
models of the nucleosynthesis through the cosmic ray spallation were
motivated to simultaneously agree with whole the observational Li-Be-B
abundances~\cite{li6metal,li6LSTC,li6FO}.  On the other hand, recently
it was claimed that the observational $^6$Li abundance in halo stars
is too abundant from the point of view of the cosmic ray energy if
$^9$Be is fit by the model of the cosmic-ray metal~\cite{ramaty}.
Therefore, there seems to be some uncertainties in the models of the
cosmic ray spallation.  In this situation, however, at least it would
be safe to assume that $^6$Li abundance increases as the metallicity
increases.  Today we observe only the $^6$Li to $^7$Li ratio in
low-metallicity ([Fe/H] $\leq - 2.0$) halo stars~\cite{li6_obs},
\begin{equation}
      \label{eq:obs6}
      r_{6,7,halo}^{obs} \equiv (\lisix/\liseven)^{obs}_{halo}
      = 0.05 \pm 0.02.
\end{equation}
We take this value as an upper bound on the primordial value
$r_{6,7}^{obs}$,
\begin{equation}
      \label{eq:obs6_upp}
      r_{6,7}^{obs} \le r_{6,7,halo}^{obs}.
\end{equation}

\section{Photodissociation and non-thermal production of {\bf $^6$Li}}
\label{sec:photodis}

\subsection{Photodissociation}
\label{subsec:overview}

In order to discuss the effect of high-energy photons on BBN, we need
the shape of the photon spectrum induced by the primary high-energy
photons from the decay of the massive particle $X$. In the thermal
bath (mixture of photons $\gamma_{\rm BG}$, electrons $e^{-}_{\rm
BG}$, and nucleons $N_{\rm BG}$), high energy photons lose their
energy by the following cascade processes:
   \begin{itemize}
    \item Double-photon pair creation
($\gamma +\gamma_{\rm BG} \rightarrow e^{+} +e^{-}$)
    \item Photon-photon scattering
($\gamma +\gamma_{\rm BG} \rightarrow \gamma +\gamma$)
    \item Pair creation in nuclei
($\gamma  +N_{\rm BG} \rightarrow e^{+}  +e^{-} + N$)
    \item Compton scattering
($\gamma  +e^{-}_{\rm BG} \rightarrow \gamma  +e^{-}$)
    \item Inverse Compton scattering
($e^{\pm} +\gamma_{\rm BG} \rightarrow e^{\pm}  +\gamma$)
\end{itemize}
In this study we numerically solved the Boltzmann equation including
the above processes, and obtained the distribution function of photons,
$f_\gamma (E_\gamma)$.

The cascade photons induce the photodissociation of the light
elements, which modifies the result of standard BBN (SBBN). The
evolutions of the light nuclei abundances are governed by the
following Boltzmann equation:
\begin{eqnarray}
    \frac{dn_N}{dt} + 3Hn_N &=& \left[\frac{dn_N}{dt}\right]_{\rm SBBN}
    - n_N \sum_{N'}\int dE_\gamma
    \sigma_{N\gamma\rightarrow N'}(E_\gamma) f_\gamma (E_\gamma)
   \nonumber \\ &&
    + \sum_{N''}n_{N''} \int dE_\gamma
    \sigma_{N''\gamma\rightarrow N} (E_\gamma) f_\gamma (E_\gamma),
\end{eqnarray}
where $n_N$ is the number density of the nuclei $N$, and
$[dn_N/dt]_{\rm SBBN}$ denotes the SBBN contribution to the Boltzmann
equation.  In Table~\ref{table:pf}, we list the photodissociation
processes included in our computation.  In this study the model
parameters are the baryon to photon ratio ($\eta$), the lifetime of
$X$ ($\tau_X$), the mass of $X$ ($m_X$), and the yield variable
$Y_{X}$ of $X$ after electron-positron annihilation,
\begin{equation}
      \label{yx}
   Y_X = n_X/n_{\gamma},
\end{equation}
where $n_{\gamma}$ is the number density of photon.\footnote{
Note that in the reference~\cite{HKKM}, $Y_X = n_X/n_{\gamma}$ is
defined before electron-positron annihilation ($e^+e^-$ ann.).  Then
they have a relationship $Y_X|_{\rm after \ e^+e^- ann.} = \frac4{11}
Y_X|_{\rm before \ e^+e^- ann.}$.}
In this paper we assume that $X$ decays only into photons, i.e.,
$m_XY_X$ corresponds to $\Delta \rho_{\gamma}/n_{\gamma}$.  Then, the
photodissociation rates depend on the combination $m_XY_X$ which
characterizes the amount of the energy of the injected photons
$\Delta\rho_{\gamma}$ as far as $m_X$ is much larger than 20
MeV~\cite{KMapj}.

\subsection{Non-thermal $^6$Li production}
\label{subsec:nt_li6}

As pointed out by Jedamzik~\cite{jedamzik}, both $\T$ and $\hethree$
are produced through the photodissociation of $\hefour$,
\begin{eqnarray}
      \label{eq:he4-gamma}
      ^4{\rm He} + \gamma \longrightarrow \left\{
        \begin{array}{ll}
             n + \hethree \\
             p + \T
        \end{array}
      \right.
\end{eqnarray}
They are still energetic and have kinetic energies enough to produce
$^6$Li through the following processes with the background $\hefour$:
\begin{eqnarray}
     \T + \hefour &\longrightarrow& \lisix + n,
     \label{eq:li6_T} \\
     \hethree + \hefour &\longrightarrow& \lisix + p,
     \label{eq:li6_he3}
\end{eqnarray}
until they are stopped by the ionization loss through the plasma
excitation in the electromagnetic plasma. The threshold energy of the
$^6$Li production is $E_{6 \hethree}^{th} = 4.03\mev$ for $\hethree$,
and $E_{6 \T}^{th} = 4.80 \mev$ for T. Then, the abundance of $^6$Li
produced through the non-thermal (NT) process in Eq.~(\ref{eq:li6_T})
is governed by
\begin{eqnarray}
      \label{eq:dn6_dt}
      \lefteqn{\left[\frac{dn_{\lisix}}{dt}\right]_{\rm NT} =} \nonumber
      \\ && n_{\hefour} \int^{\infty}_{E_4^{th}+4E_6^{th}} dE_{\gamma}
      \sigma_{\hefour(\gamma,p)\T}(E_{\gamma}) f_{\gamma}(E_{\gamma})
      \int_{E_6^{th}}^{(E_{\gamma}-E_4^{th})/4}n_{\hefour}
      \sigma_{\T(\alpha,n)\lisix}(E) \left(\frac{dE}{dx}\right)^{-1} dE,
\end{eqnarray}
where $n_{\lisix} ( n_{\hefour} )$ denotes the number density of
$\lisix ( \hefour )$. $\sigma_{\hefour(\gamma,p)\T}(E_{\gamma})$ is
the cross section of the $\hefour$ photodissociation, $E_4^{th}$ is
the threshold energy of the photodissociation process,
$f_{\gamma}(E_{\gamma})$ is the photon spectrum which are obtained by
solving the Boltzmann equation, $\sigma_{\T(\alpha,n)\lisix}(E)$ is
the cross section of the process in Eq.~(\ref{eq:li6_T}). $dE/dx$
denotes the rate of the ionization loss while the charged particle
$\T$ is running a distance $dx$ in the electromagnetic plasma. The
rate of the ionization loss is expressed by~\cite{jackson}
\begin{equation}
      \label{eq:ion_loss}
      \frac{dE}{dx} =
      \frac{Z^2 \alpha}{\beta^2} \omega_p^2 \ln\left(\frac{\Lambda m_e
        \beta^2}{\omega_p}\right),
\end{equation}
where $\omega_p^2$ is plasma frequency (= $4\pi n_e \alpha /m_e$),
$n_e$ is the electron number density, $m_e$ is electron mass, $Z$ is
the charge, $\Lambda \sim \order{(1)}$ is a constant and $\beta$ is
the velocity.  The effect of the process (\ref{eq:li6_he3}) is
described by replacing the suffix $\T$ with $\hethree$ in
Eq.~(\ref{eq:dn6_dt}).

We include the above two processes of the non-thermal $^6$Li
production in BBN code and compute the $^6$Li abundance. In the
computation we adopt the experimental cross section
$\sigma_{\T(\alpha,n)\lisix} = 35 \pm 1.4$ mb~\cite{koepke_brown}
commonly for the two processes. Because we have only one data point at
the kinetic energy $E_T$ = 28 MeV in the laboratory system, we assume
that the cross section is constant for whole the energy region and
neglect its energy dependence. Then, we integrate the second factor in
Eq.~(\ref{eq:dn6_dt}) up to a high energy. One can easily find that
there exists a serious problem in this procedure if it is compared to
the case of the original photodissociation where the photodissociation
rates steeply decrease as the energy increases.  Because we have the
experimental data for the $\hefour$ photodissociation rates only up to
about 100 MeV for the photon
energy~\cite{CJP53-802,PLB47-433,SJNP19-598}, we should interpolate
the photodissociation rates in a high energy region because of the
mild dumping of the integrand. Then, the integration has a large
uncertainty ($\sim$ 20 $\%$) when we change the upper limit of the
integration from 500 MeV to 1 GeV.\footnote{
In addition, there may be another larger uncertainty which comes from
the differences of the method for the interpolation because we do not
know the correct shape of the cross sections. In this case, the
obtained constraint would be weaker.}  
Therefore, in this situation we adopt 20$\%$ errors for the
non-thermal $^6$Li production rates and
perform the Monte Carlo simulation including them.\footnote{
If the cross section $\sigma_{\T(\alpha,n)\lisix}$ decreases at high
energy like other nuclear interactions, the $^6$Li production is less
important. As shown later, the resultant constraint is not changed
even if we neglect the $^6$Li production. } 

\subsection{Constraint from cosmic microwave background}

In addition to the photodissociation process, there also exists an
another constraint. A radiative decay process releases a net photon
energy into the electromagnetic plasma. The emitted photons should be
thermalized soon, otherwise the photon spectrum deviates from the
blackbody, which contradicts the observation of the cosmic microwave
background (CMB)~\cite{fixsen}. This leads to the following
constraints:
\begin{equation}
      \label{eq:mu_dis}
      m_X Y_X \lesssim 2.0 \times 10^{-12} \gev \left
      ( \frac{\tau_X}{10^{10} \sec} \right)^{\frac12},
\end{equation}
for $\mu$-distortion ($1.8 \times 10^{6} \sec \left(\Omega_B h^2/0.02
\right)^\frac23 \lesssim \tau_X \lesssim 2.3 \times 10^9 \sec
\left(\Omega_B h^2/0.02 \right)$), and
\begin{equation}
      \label{eq:y_dis}
      m_X Y_X \lesssim 1.9 \times 10^{-12} \gev \left
      ( \frac{\tau_X}{10^{10} \sec} \right)^{\frac12},
\end{equation}
for $y$-distortion ($2.3 \times 10^9 \sec \left(\Omega_B h^2/0.02
\right) \lesssim \tau_X \lesssim 10^{12} \sec$).

\section{Comparison with observational light element abundances}
\label{sec:obs}
\begin{figure}
      \begin{center}
          \centerline{\psfig{figure=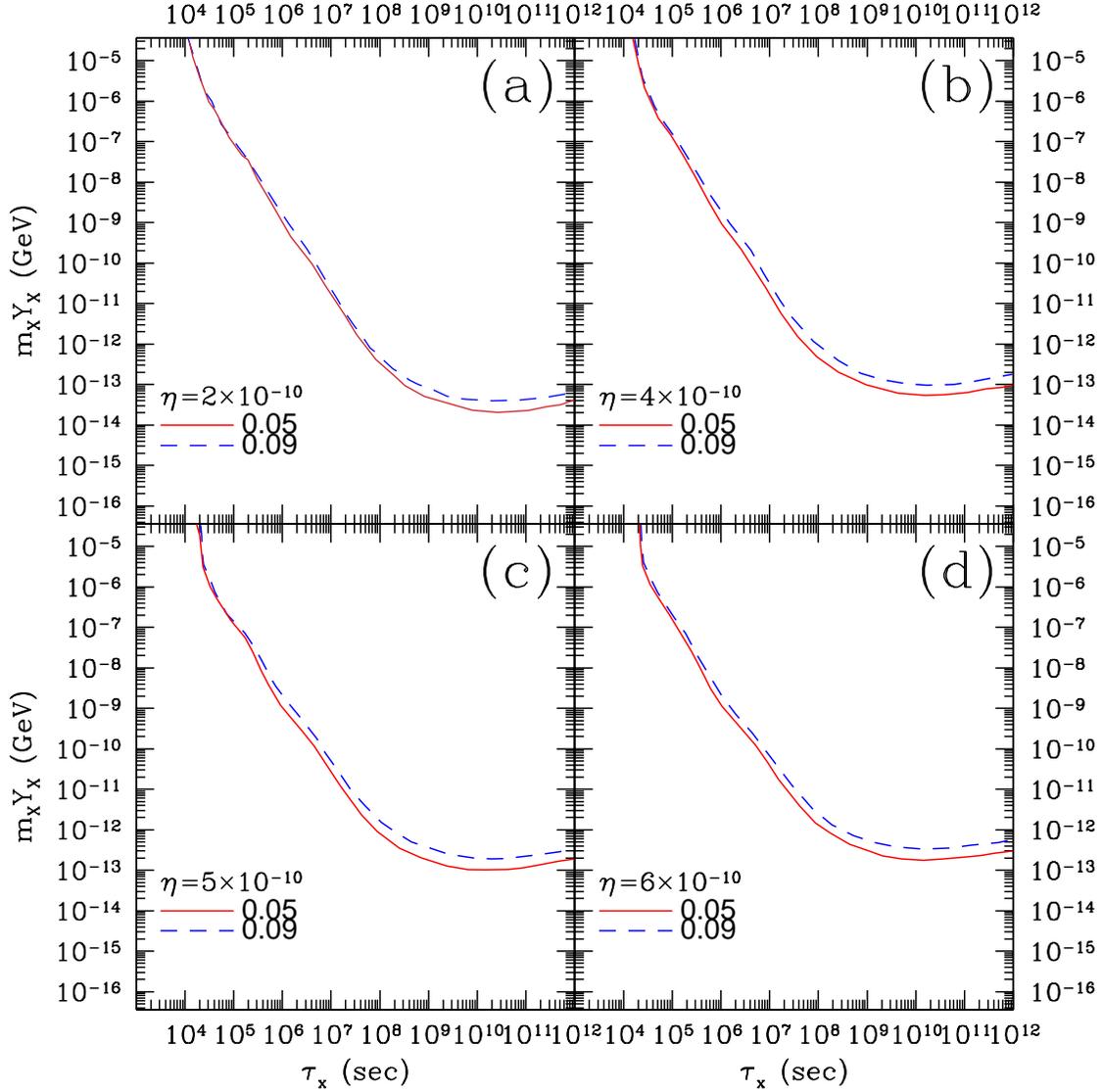,width=17cm}}
          \caption{Plot of
          $^6$Li to $^7$Li ratio in ($\tau_X$, $m_X Y_X$) plane for
          various baryon to photon ratio ($\eta = n_B/n_{\gamma}$) in
          (a) $\eta = 2 \times 10^{-10}$, (b) $\eta = 4 \times 10^{-10}$,
          (c) $\eta = 5 \times 10^{-10}$, and (d) $\eta = 6 \times
          10^{-10}$. The solid line denotes the observational mean value
          of $^6$Li / $^7$Li and the dashed line denotes the
          observational 2-$\sigma$ upper bound.}
          \label{fig:pn67}
      \end{center}
\end{figure}

In Fig.~\ref{fig:pn67} we plot the theoretically predicted $^6$Li to
$^7$Li ratio ($\equiv r_{6,7}^{th}$) in ($\tau_X$, $m_X Y_X$) plane.
The solid line represents the model parameters which predict the
observational mean value of $r_{6,7}^{th}$ and the dashed line denotes
the observational 2-$\sigma$ upper bound.  From the figure, one may
think that the mean value of the theoretical prediction constrains
$m_X Y_X$ severely.  We should bear in mind, however, that the
theoretical prediction has a large uncertainty which comes from the
errors of the production rates, and in addition the observational
constraint also has a large error.  To take account of these
uncertainties systematically, we performed the Maximum Likelihood
analysis~\cite{HKKM} including both the theoretical and the
observational errors.  Here we assume that the theoretical predictions
of (D/H)$^{th}$, $Y^{th}$, ${\rm log_{10}}[(\liseven/{\rm H})^{th}]$,
$r_{3,2}^{th} = (\hethree/\D)^{th}$, and $r_{6,7}^{th}$ obey the
Gaussian probability distribution functions (p.d.f.'s) with the widths
given by the 1-$\sigma$ errors.  Concerning the observational values,
(D/H)$^{obs}$, $Y^{obs}$, and ${\rm log_{10}}[(\liseven/{\rm
H})^{obs}]$ are assumed to obey the Gaussian p.d.f.'s while we treat
$r_{3,2}^{obs}$ and $r_{6,7}^{obs}$ as non-Gaussian variables
\cite{HKKM}.

\begin{figure}
      \begin{center}
       \centerline{\psfig{figure=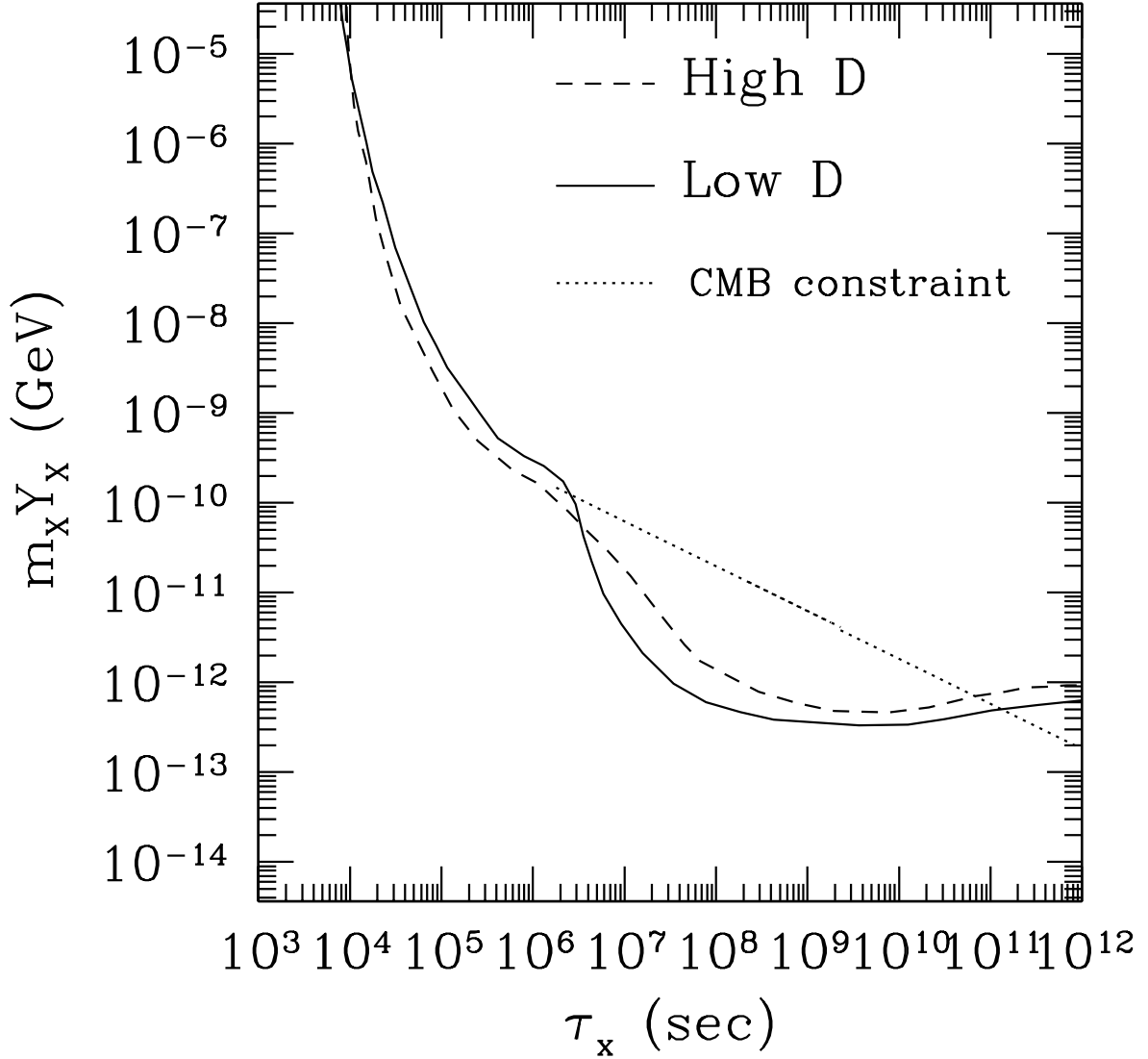,width=17cm}}
          \caption{Plot of the contour of the confidence level in
          ($\tau_x, m_X Y_X$) plane. The solid (dashed) line denotes the
          95$\%$ C.L. for Low D (High D) projected on $\eta$ axis.  The
          dotted line denotes the upper bound which comes from CMB
          constraint.}
          \label{fig:mxyx}
      \end{center}
\end{figure}

In Fig.~\ref{fig:mxyx} we plot the results of the $\chi^2$ fitting by
using the method of the Maximum Likelihood analysis.  The solid
(dashed) line denotes the Low D (High D) constraint. The dotted line
denotes the upper bound from the CMB constraint.  In the figure, the
region below the lines are consistent with the observations.  The
constraint from the CMB is almost always weaker than that from
BBN. The main feature of the difference between High D and Low D is
that the Low D constraint is severer than High D for a relatively long
lifetime case ($\tau_X \gtrsim 3 \times 10^6 \sec$). That is because
High D constraint modestly allows the overproduction of $\hethree$
accompanying the $\hefour$ photodissociation. On the other hand, the
High D constraint is more stringent for shorter lifetimes since the D
dissociation is more important than the $\hefour$ photodissociation.

The obtained upper bound does not change our earlier results so much
\cite{HKKM}. It became slightly weaker because we included the $\eta$
dependence for the photodissociation rates
($\Gamma \propto 1/\eta$) in this analysis.\footnote{
The $\eta$ dependence is understood as follows.  The soft photons
produced in the electromagnetic cascade scatter off the background
electrons and nucleons and lose their energy.  Thus the number density
of soft photons with energy larger than the threshold decreases as
scattering rate which is proportional to $\eta$. Therefore, the
photodissociation rates are proportional to $1/\eta$.}
We  find that the non-thermally produced $^6$Li mildly
contributes to the bound.\footnote{
Tritium is unstable with the lifetime $\tau_{\T} = 5.614 \times 10^8$
sec, and decays into $\hethree$ whose charge is two.  Thus, because
$\hethree$ subjects to stop easier than $\T$ by the ionization loss,
we might overestimate the $\lisix$ production in parameter regions
where the stopping time $\tau_{stop} = \int^0_E (dE/dt)^{-1}dE \simeq
2.5\times 10^9 \sec (T/\ev)^{-3}(E/\mev)^{3/2}$ is longer than the
lifetime of Tritium, i.e., for $T \lesssim 1.7 \ev$.  Therefore, at a
long lifetime $\tau_X \gtrsim 5 \times 10^{11} \sec$, our constraint
might become weaker by about factor two.  However, it is
expected that the effect would not change the result significantly
because the $\hethree$ overproduction gives a severer constraint
there.}
The main reason is that both the theoretical computation and
observational data have very large uncertainties which amount to about
30 -- 40 $\%$.

Assuming that the parent massive particle is the gravitino and that it
dominantly decays into a photon and a photino ($\psi_{3/2} \to
\tilde{\gamma} + \gamma$), the lifetime $\tau_{3/2}$ is related to the
gravitino mass $m_{3/2}$ as
\begin{eqnarray}
    \tau_{3/2}\simeq 4\times 10^5{\rm ~sec} \times
    (m_{3/2}/1 {\rm ~TeV})^{-3}.
\end{eqnarray}

\begin{figure}
      \begin{center}
          \centerline{\psfig{figure=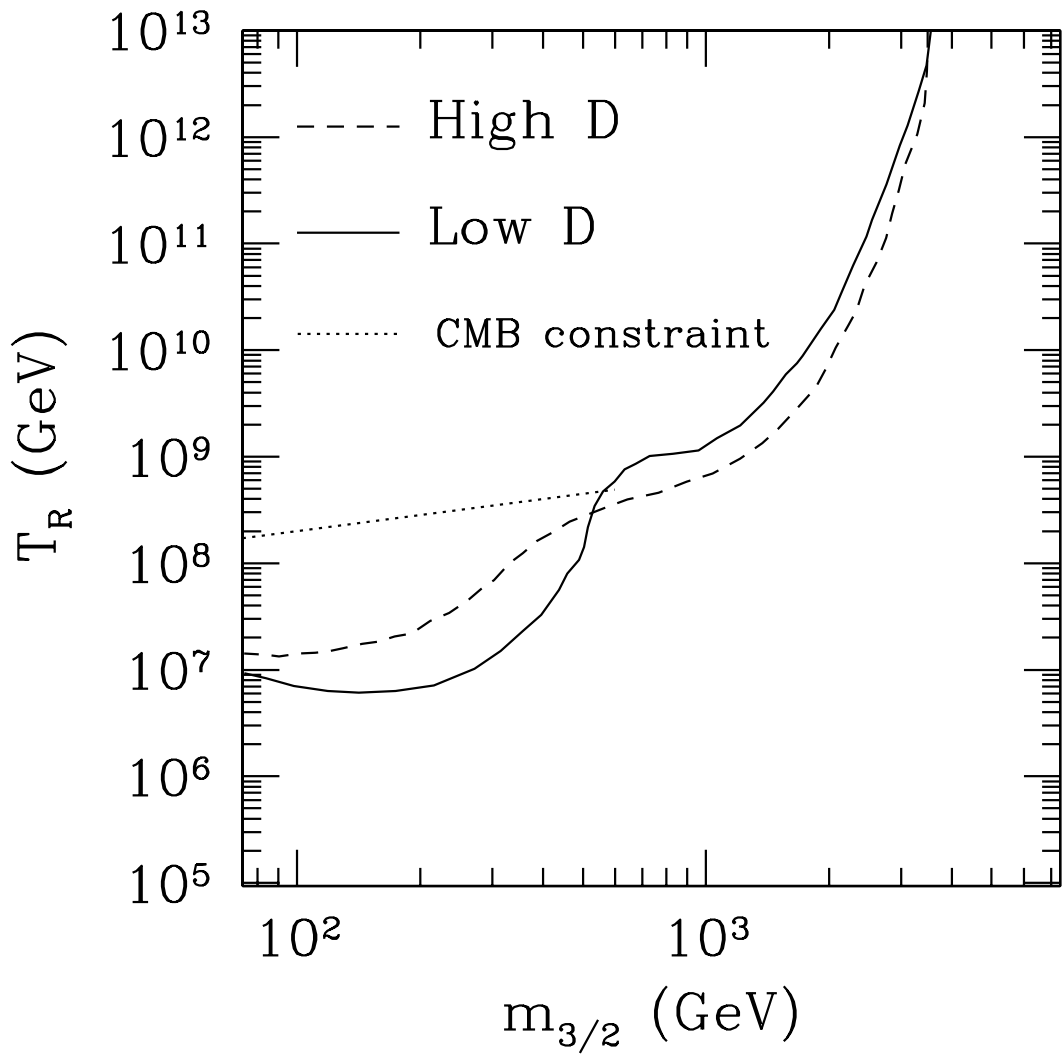,width=17cm}}
          \caption{Plot of the contour of the confidence level in
          ($m_{3/2}$, $T_R$) plane. The solid (dashed) line denotes the
          95$\%$ C.L. for Low D (High D).  The dotted line denotes the
          upper bound which comes from CMB constraint.}
          \label{fig:rht}
      \end{center}
\end{figure}

Assuming that the gravitino is produced through the thermal scattering
in the reheating process after
inflation,\footnote{
Although these days it was claimed that gravitinos are also produced
in the preheating epoch non-thermally~\cite{KKLP,GRT,Lyth99}, we do
not consider such processes here because there are some ambiguities on
the estimations and they depend on various model parameters.  If the
non-thermal production is effective, however, the obtained constraint
might be severer.}
we relate the abundance $Y_{3/2} = n_{3/2}/n_\gamma$ of the gravitino
with the reheating temperature $T_R$~\cite{KM},
\begin{eqnarray}
    Y_{3/2} \simeq 1.1 \times 10^{-11} \times (T_R/10^{10}{\rm GeV}).
\end{eqnarray}
In Fig.~\ref{fig:rht} we plot the upper bound on the reheating
temperature after inflation at 95$\%$ C.L. as a function of the
gravitino mass. Here we can read off the constraint by using the
relationship of the scaling, $\Delta \rho_{\gamma}/n_{\gamma} =
\frac12 m_{3/2}Y_{3/2} (= m_X Y_X)$ because we assumed X decays into
two photons.  From the figure we can obtain the upper bound on the
reheating temperature:
\begin{eqnarray}
   m_{3/2}=100{\rm ~GeV} ~~~
    (\tau_{3/2}\simeq 4\times 10^8{\rm ~sec}) &:&
    T_R \lesssim 1 \times 10^7 {\rm ~GeV},
   \nonumber \\
    m_{3/2}=1{\rm ~TeV} ~~~
    (\tau_{3/2}\simeq 4\times 10^5{\rm ~sec}) &:&
    T_R \lesssim 1 \times 10^9 {\rm ~GeV},
   \nonumber \\
    m_{3/2}=3{\rm ~TeV} ~~~
    (\tau_{3/2}\simeq 1\times 10^4{\rm ~sec}) &:&
    T_R \lesssim 9\times 10^{11} {\rm ~GeV},
    \label{eq:eq:tr_mass}
\end{eqnarray}
at 95$\%$ C.L.

\section{Conclusion}
\label{sec:conclusion}

In this paper we have studied the effects on primordial
nucleosynthesis of the radiative decay of a long-lived massive
particle $X$ using new observational data.  We have also considered
the non-thermal $\lisix$ production caused by energetic T and
$\hethree$ produced by the $\hefour$ photodissociation.  We obtained
the photon spectrum through the electromagnetic cascade process by
solving a set of Boltzmann equations numerically.  In addition, to
estimate the theoretical errors we performed Monte Carlo simulation
including the theoretical uncertainties which come from those of
nuclear reaction rates.  To obtain the degree of agreements between
theory and observation, we performed the Maximum Likelihood method and
the $\chi^2$ fitting including both the observational and theoretical
errors.

As a result we have obtained the upper bound on the abundance $m_XY_X$
as a function of its lifetime $\tau_X$.  The result does not change
our previous works significantly.  This is because the theoretical and
observational errors for $\lisix$ are significantly large, and it
contributes to the constraints more weakly than the $\hethree$
overproduction accompanying the $\hefour$
photodissociation. Therefore, we have concluded that it is premature
to emphasize the importance of the non-thermal production of $^6$Li.

We have also applied the results obtained by a generic radiatively
decaying particle to gravitino $\psi_{3/2}$, and we have got the upper
bound on the reheating temperature after primordial inflation as a
function of the mass, $T_R \lesssim 10^7 - 10^9 \  \gev$ for $m_{3/2} =
100 \ \gev - 1 \ \tev$ (95 $\%$ C.L.).

\section{Acknowledgments}

The work of M.K. is supported by Priority Area 707 ``Supersymmetry and
Unified Theory of Elementary Particles."  The work of T.M. is
supported by the Grant-in-Aid for Scientific Research from the
Ministry of Education, Science, Sports and Culture of Japan No.\
12047201.


\begin{table}[t]
\begin{center}
\begin{tabular}{rlrrr}
& {Photodissociation Reactions} & 1-$\sigma$ Uncertainty & Threshold Energy
& Ref.\\
\hline
   1. &   ${\rm D} + \gamma \rightarrow p + n$
                        &  6\% &  2.2 MeV & \cite{Evans}\\
   2. &   ${\rm T} + \gamma \rightarrow n + {\rm D}$
                        & 14\% &  6.3 MeV & \cite{ZP208-129,PRL44-129}\\
   3. &   ${\rm T} + \gamma \rightarrow p + 2n$
                        &  7\% &  8.5 MeV & \cite{PRL44-129} \\
   4. &$^3{\rm He} + \gamma \rightarrow p + {\rm D}$
                        & 10\% &  5.5 MeV & \cite{PL11-137} \\
   5. &$^3{\rm He} + \gamma \rightarrow n + 2p
                      $ & 15\% &  7.7 MeV & \cite{PL11-137} \\
   6. &$^4{\rm He} + \gamma \rightarrow p + {\rm T}$
                        &  4\% & 19.8 MeV & \cite{PL11-137} \\
   7. &$^4{\rm He} + \gamma \rightarrow n +~^3{\rm He}$
                        &  5\% & 20.6 MeV & \cite{CJP53-802,PLB47-433} \\
   8. &$^4{\rm He} + \gamma \rightarrow p + n + {\rm D}$
                        & 14\% & 26.1 MeV & \cite{SJNP19-598} \\
   9. &$^6{\rm Li} + \gamma \rightarrow {\rm anything}$
                        &  4\% &  5.7 MeV & \cite{Berman} \\
10. &$^7{\rm Li} + \gamma \rightarrow 2n + {\rm anything}$
                        &  9\% & 10.9 MeV & \cite{Berman} \\
11. &$^7{\rm Li} + \gamma \rightarrow n +~^6{\rm Li}$
                        &  4\% &  7.2 MeV & \cite{Berman} \\
12. &$^7{\rm Li} + \gamma \rightarrow~^4{\rm He} + {\rm anything}$
                        &  9\% &  2.5 MeV & \cite{Berman} \\
13. &$^7{\rm Be} + \gamma \rightarrow p +~^6{\rm Li}$
                        &  4\% &          & \\
14. &$^7{\rm Be} + \gamma \rightarrow~{\rm anything~except}~^6{\rm Li}$
                        & 9\%  &          & \\
\end{tabular}
\caption{List of photodissociation processes, and the 1-$\sigma$ uncertainty
in the cross sections.  Since there is no experimental data on
photodissociation of $^7$Be, we assume that the rate, threshold, and
uncertainty for Reaction 13 is the same as for Reaction 11, and the
rate for Reaction 14 is the sum of the rates for Reactions 10 and 12.}
\label{table:pf}
\end{center}
\end{table}

\newpage

\newpage

\end{document}